\documentclass{article}
\usepackage[utf8]{inputenc}
\usepackage{listings}
\usepackage{xcolor}
\usepackage{qcircuit}
\usepackage{subfigure}
\usepackage{braket}
\usepackage{amsmath,bm}
\usepackage[hyphens]{url}
\usepackage{hyperref}
\usepackage[hyphenbreaks]{breakurl}
\usepackage{graphicx}
\usepackage{authblk}
\usepackage{booktabs}

\usepackage[
backend=biber,
style=alphabetic,
]{biblatex}
\newtheorem{definition}{Definition}
\definecolor{bgray}{RGB}{245,245,244}
\newcommand{\tcode}[1]{\colorbox{bgray}{\texttt{#1}}}

\begin{document}
\title{VeriQBench: A Benchmark for Multiple Types of Quantum Circuits}
\author[1,3]{Kean Chen}
\author[1,3]{Wang Fang}
\author[1,\thanks{guanj@ios.ac.cn}]{Ji Guan}
\author[4]{
Xin Hong}
\author[1,3]{Mingyu Huang}
\author[1,3]{Junyi Liu}
\author[2]{Qisheng Wang}
\author[1,2,\thanks{yingms@ios.ac.cn}]{Mingsheng Ying}
\affil[1]{State Key Laboratory of Computer Science, Institute of Software, Chinese Academy of Sciences, China}
\affil[2]{Department of Computer Science and Technology, Tsinghua University, China}
\affil[3]{The University of Chinese Academy of Sciences, China}
\affil[4]{Centre for Quantum Software and Information, University of Technology Sydney, Australia}
\date{\today}
\maketitle
\begin{abstract}

In this paper, we introduce \emph{VeriQBench} --- an open source benchmark for quantum circuits. It offers high-level quantum circuit abstractions of various circuit types,  including: 1) combinational, 2) dynamic, 3) sequential, and 4) variational quantum circuits, which cover almost all existing types of quantum circuits in the literature.  
Meanwhile, \emph{VeriQBench} is a versatile benchmark which can be used in verifying quantum software for different applications, as is evidenced by the existing works including quantum circuit verification (e.g., equivalence checking~\cite{hong2021approximate,WLY21} and model checking~\cite{ying2021model}), simulation (e.g., fault simulation), testing (e.g., test pattern generation~\cite{chen2022automatic}) and debugging (e.g., runtime assertions~\cite{li2020projection}). 
All the circuits are described in OpenQASM and are validated on Qiskit and QCOR simulators. With the hope that it can be used by other researchers,  \emph{VeriQBench} is released at: \url{https://github.com/Veri-Q/Benchmark}.
\end{abstract}

\newpage

\tableofcontents
\section{Introduction}
Quantum circuits are the basic  components in quantum computing.
With the rapid development in hardware implementation of quantum circuits in  experiments, there also emerge many approaches to characterise and evaluate  quantum computing models by  simulating and comparing quantum circuits, such as quantum noise effects~\cite{liu2021benchmarking}, variational circuit training~\cite{broughton2020tensorflow} and  (approximate) equivalence checking 
~\cite{viamontes2007checking,yamashita2010fast,burgholzer2020improved,burgholzer2020advanced,hong2021approximate}. In particular,
quantifying the practical performance of some carefully chosen quantum circuits is a common way to evaluate the performance of quantum devices, especially in the current NISQ (Noisy Intermediate-Scale Quantum) era, where quantum noise from the surrounding environment is unavoidable.  These quantum circuits are generated from different fields, such as random protocols~\cite{arute2019quantum},
quantum algorithms~\cite{nielsen2002quantum,childs2017lecture}, 
including variational quantum algorithms~\cite{peruzzo2014variational,farhi2014quantum,hadfield2019quantum,tilly2021variational,cerezo2021variational} and classical reversible circuits~\cite{abdessaied2016reversible}. Subsequently, in the recent years, diverse quantum circuit benchmarks have been  proposed with different evaluation metrics, such as QASMBench~\cite{li2020qasmbench}, quantum LINPACK benchmark~\cite{dong2021random},  SupermarQ~\cite{tomesh2022supermarq} and the quantum volume protocol~\cite{cross2019validating}. 

In this paper, we present a new quantum circuit benchmark, called \emph{VeriQBench}, for a different purpose. It was initially designed for testing our own QDA (Design Automation for Quantum Computing) tool series $\textit{VeriQ}$. But  we hope it can be  applicable in various fields, given the following features:
\begin{itemize}
    \item {\bf Diversity:} It includes the most commonly-used  quantum circuits, namely  combinational quantum circuits, dynamic quantum circuits, sequential quantum circuits, and variational quantum circuits.
    These circuits cover all types of quantum circuits existing in the literature. Furthermore, the quantum circuits in our benchmark are diverse in structure (e.g. the layout of 2-qubit circuits) and complexity (e.g. circuit size and depth).
    \item {\bf Scalability:} The circuit scales in our benchmark vary widely, ranging from 2 qubits up to $>50$ qubits, which are divided into three classes, \textit{i.e.} small-scale ($<20$ qubits), medium-scale ($20-50$ qubits) and large-scale ($>50$ qubits).
    \item {\bf Easy-to-use:} All the circuits in our benchmark are described using the OpenQASM quantum assembly language and provided as ``.qasm'' files. Furthermore, most of them can be converted to other representations such as Q\#, PyQuil, Cirq, etc. through \emph{q-convert} tool, which is available online: \url{http://quantum-circuit.com/qconvert}.
    \item {\bf Evolvement:}  For most of the quantum circuits in our benchmark, we provide a series of scripts for users to generate quantum circuits of an arbitrary number of qubits. This ensures that our benchmark can  evolve as the available circuit scales of quantum technologies increase.
\end{itemize}

Our benchmark can work with different evaluation metrics to verify quantum software --- comparing and assessing the effectiveness and efficiency of quantum and classical (simulation) algorithms designed for quantum circuits. The circuits in our benchmark are collected from researches in a variety of fields, including equivalence checking~\cite{hong2021approximate,WLY21}, circuit testing~\cite{chen2022automatic} and circuit optimizing~\cite{amy2014polynomial,nam2018automated}. This improves  the diversity and practical feasibility of the benchmark. On this basis, the circuit types and scales are further extended and standardized to form a scalable and evolvable benchmark.
To demonstrate the validity, all quantum circuits in our benchmark are implemented and validated by Qiskit~\cite{qiskit} --- an open source SDK for quantum computation, and QCOR~\cite{qcor} --- a programming language and a compiler for the heterogeneous quantum-classical model of computation.


This paper is organised as follows: we start from the basic type of quantum circuits, namely combinational quantum circuits in Section~\ref{sec:combinational}, which includes some fundamental quantum algorithms, reversible circuits, qubit mapping and random quantum circuits. Then dynamic, sequential and variational quantum circuits are presented in Sections~\ref{sec:dynamic},~\ref{sec:sequential} and~\ref{sec:variational}, respectively. In Section~\ref{section_val}, we validate all quantum circuits on Qiskit and QCOR. In the last section, we discuss some example applications of the benchmarks. Of course, we hope that this benchmark can be used by other researchers. The benchmark \emph{VeriQBench} is released at: \url{https://github.com/Veri-Q/Benchmark}.

\section{Combinational Quantum Circuits}\label{sec:combinational}


The most basic and commonly used type of quantum circuit is the combinational quantum circuit, with examples  including some of the popular quantum algorithms such as the Bernstein-Vazirani algorithm, quantum Fourier transform and quantum phase estimation. In our benchmark, we include the most commonly used combinational circuits and divide them into four categories: quantum algorithms, reversible circuits, circuits for qubit mapping, and random quantum circuits.


\subsection{Quantum Algorithms}

\subsubsection{Bernstein-Vazirani Algorithm}

\textbf{Description}. 
Bernstein-Vazirani algorithm \cite{bernstein1997quantum} is an algorithm that can be used to find the hidden string $s$ given a boolean function $f(x)$, where $f(x)=\langle s,x \rangle=s_0x_0\oplus s_1x_1\oplus\cdots \oplus s_nx_n$. For the classical algorithm, it normally needs $\mathcal{O}(n)$ times to complete this task using a bit-by-bit inquiring method. But, it only needs $\mathcal{O}(1)$ times using the Bernstein-Vazirani algorithm suppose that you are given an orcle $O_s$, where $O_s\ket{x}\ket{y}=\ket{x}\ket{f(x)\oplus y}$.

First, apply a series of Hadamard gates to state $\ket{0}\cdots\ket{0}\ket{1}$ to change the state to $\frac{1}{\sqrt{2}^n} \sum_{x=0}^{2^n-1}{\ket{x}}\otimes \frac{\ket{0}-\ket{1}}{\sqrt{2}}$. Then, applying the oracle, the state will become $\frac{1}{\sqrt{2}^n} \sum_{x=0}^{2^n-1}{(-1)^{\langle s,x \rangle}\ket{x}}\otimes \frac{\ket{0}-\ket{1}}{\sqrt{2}}$. Finally, apply a series of Hadamard gates, and the final state will change to $\ket{s}\ket{1}$. At the end, measuring the first $n$ qubits gives the value of $s$.

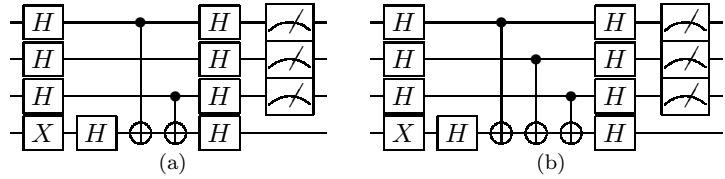
\begin{figure}[h]
\centerline{
\subfigure[]{
\centering
\Qcircuit @C=.5em @R=0em @!R {
& \gate{H} & \qw      &\ctrl{3} &\qw     & \gate{H} &\qw & \meter & \qw &&\\
& \gate{H} & \qw      &\qw      &\qw     & \gate{H} &\qw & \meter & \qw &&\\
& \gate{H} & \qw      &\qw      &\ctrl{1}& \gate{H} &\qw & \meter & \qw &&\\
& \gate{X} & \gate{H} &\targ    &\targ   & \gate{H} &\qw & \qw    & \qw &&\\
}
}
\subfigure[]{
\centering
\Qcircuit @C=.5em @R=0em @!R {
& \gate{H} & \qw      &\ctrl{3} &\qw &\qw     & \gate{H} &\qw & \meter & \qw &&\\
& \gate{H} & \qw      &\qw      &\ctrl{2} &\qw     & \gate{H} &\qw & \meter & \qw &&\\
& \gate{H} & \qw      &\qw      &\qw &\ctrl{1}& \gate{H} &\qw & \meter & \qw &&\\
& \gate{X} & \gate{H} &\targ    &\targ &\targ   & \gate{H} &\qw & \qw    & \qw &&\\
}
}
}
\caption{Two examples of the Bernstein-Vazirani algorithm, where (a) $s=101$ and (b) $s=111$.}
\label{cir_bv_alg}
\end{figure}

For $s=101$ and $s=111$, the corresponding circuits are shown in Figure \ref{cir_bv_alg}. In these two circuits, an $X$ gate is added at the last qubit to change the initial state from $\ket{0}$ to $\ket{1}$, and the $CNOT$ gates are used for implementing the oracle. For these circuits, setting input state $\ket{0}\cdots \ket{0}$ and measuring at the end will give the hidden string $s$.

\noindent\textbf{OpenQASM Code}. The following gives the description of the circuit for $s=101$ using OpenQASM 2.0. 
\begin{lstlisting}[]
OPENQASM 2.0;
include "qelib1.inc";
qreg q[4];
creg c[4];
h q[0];
h q[1];
h q[2];
x q[3];
h q[3];
cx q[0],q[3];
cx q[2],q[3];
h q[0];
h q[1];
h q[2];
h q[3];
\end{lstlisting}

\noindent\textbf{Generation Script}. The circuit for arbitrary hidden string $s$ can be generated using the following code. For this circuit, the number of qubits is expected to be one more than the length of the hidden string.

\begin{lstlisting}[language = Python]
def gen_bv(qubits, hiddenString):
    
    cir = QuantumCircuit(qubits, qubits)

    for i in range(qubits - 1):
        cir.h(i)

    cir.x(qubits - 1)
    cir.h(qubits - 1)
    hiddenString = list(hiddenString)
    for i in range(len(hiddenString)):
        if hiddenString[i] == "1":
            cir.cx(i, qubits - 1)

    for i in range(qubits):
        cir.h(i)

    return cir.qasm()
\end{lstlisting}



\subsubsection{Quantum Fourier Transform}

\textbf{Description}. Quantum Fourier transform \cite{nielsen2002quantum} is a commonly used algorithm in quantum computing. It turns an input state $\ket{j}$ to its Fourier transform.
$$QFT\ket{j}=\frac{1}{\sqrt{N}}\sum_{k=0}^{N-1}e^{2\pi ijk/N}\ket{k}.$$

\begin{figure}[h]
\centering
\centerline{
\Qcircuit @C=0.7em @R=0.4em {
\lstick{\ket{j_1}}  & \gate{H} &\ctrl{1}  &\ctrl{2}  &\qw      &\qw        &\qw      &\qw & \rstick{\ket{0}+e^{0.j_1j_2j_3}\ket{1}}\\
\lstick{\ket{j_2}}  & \qw      &\gate{R_2}&\qw       &\gate{H} &\ctrl{1}   &\qw      &\qw & \rstick{\ket{0}+e^{0.j_2j_3}\ket{1}}\\
\lstick{\ket{j_3}}  & \qw      &\qw       &\gate{R_3}&\qw      &\gate{R_2} &\gate{H} &\qw & \rstick{\ket{0}+e^{0.j_3}\ket{1}}\\
}
}
\caption{Quantum circuit for Fourier Transform.}
\label{cir_qft_alg}
\end{figure}
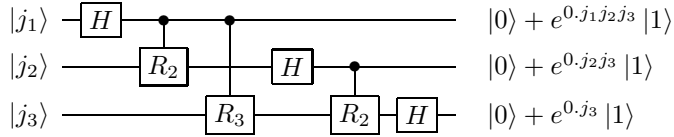

The circuit for 3-qubit quantum Fourier transform is shown in Figure \ref{cir_qft_alg}. It is easy to observe from this circuit that the quantum Fourier transform can be implemented using a series of Hadamard gates and a set of Control-$R_k$ gates, where 
$$
R_k=\begin{bmatrix} 1&0\\0&e^{2\pi i/2^k} \end{bmatrix}.
$$

\noindent\textbf{OpenQASM Code}. The OpenQASM 2.0 description of this circuit is shown below.

\begin{lstlisting}[]
OPENQASM 2.0;
include "qelib1.inc";
qreg q[3];
creg c[3];
h q[0];
cu1(pi/2) q[0],q[1];
cu1(pi/4) q[0],q[2];
h q[1];
cu1(pi/2) q[1],q[2];
h q[2];
\end{lstlisting}

\noindent\textbf{Generation Script}. The circuit for arbitrary number of qubits quantum Fourier transform can be generated using the following code. This code extended the circuit shown in Figure \ref{cir_qft_alg} to arbitrary qubits.

\begin{lstlisting}[language = Python]
def gen_qft(qubits):

    cir = QuantumCircuit(qubits, qubits)
    for q in range(qubits):
        cir.h(q)
        for tar in range(q + 1, qubits):
            theta = np.pi / 2 ** (tar - q)
            cir.cu1(theta, q, tar)

    return cir.qasm()
\end{lstlisting}

\subsubsection{Quantum Phase Estimation}
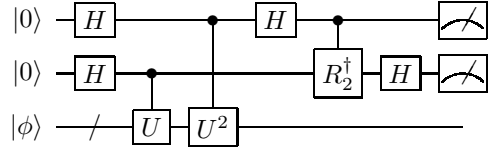
\begin{figure}[h]
\centering
\centerline{
\Qcircuit @C=0.7em @R=0.4em {
\lstick{\ket{0}}  & \gate{H} &\qw      &\ctrl{2}  &\gate{H} &\ctrl{1} &\qw &\meter \\
\lstick{\ket{0}}  & \gate{H} &\ctrl{1} &\qw       &\qw      &\gate{R_2^{\dagger}}&\gate{H}&\meter\\
 \lstick{\ket{\phi}} & {/}\qw      &\gate{U} &\gate{U^2}&\qw      &\qw&\qw&\qw\\
}
}
\caption{Quantum circuit for phase estimation.}
\label{cir_qpe_alg}
\end{figure}

\textbf{Description}. Phase estimation \cite{nielsen2002quantum} is an algorithm that can be used to estimate the phase $\varphi$ in an eigenvalue of a unitary $U$, where $U\ket{\psi}=e^{2\pi i\varphi}\ket{\psi}$ for some $\ket{\psi}$. The corresponding circuit is shown in Figure \ref{cir_qpe_alg}.

The process can be conducted by first constructing an superposition state and then applying a series of controlled-$U^{2^k}$ gates. Then, an inverse quantum Fourier transform can be used to extract the corresponding value of the phase.

{\vskip 3pt}

\noindent\textbf{OpenQASM Code}. The following gives the OpenQASM 2.0 description of this circuit. In this circuit, we suppose that $U$ is a single qubit diagonal matrix
$$U=\begin{bmatrix}1&0\\0&e^{2\pi i \varphi} \end{bmatrix},$$
and here we take $\varphi=1/1024$ as an example.

\begin{lstlisting}[]
OPENQASM 2.0;
include "qelib1.inc";
qreg q[4];
creg c[4];
h q[0];
h q[1];
h q[2];
cu1(pi/512) q[2],q[3];
cu1(pi/256) q[1],q[3];
cu1(pi/128) q[0],q[3];
h q[0];
cu1(-pi/2) q[0],q[1];
cu1(-pi/4) q[0],q[2];
h q[1];
cu1(-pi/2) q[1],q[2];
h q[2];
\end{lstlisting}

\noindent\textbf{Generation Script}. The python code for generating the circuits is as follows. You can assign the number of qubits of the circuit and also the phase $\varphi$. Then, running this circuit will give the estimation of the phase to a precision restricted by the number of qubits.

\begin{lstlisting}[language = Python]
def gen_pe(qubits,the_phase):

    cir=QuantumCircuit(qubits+1,qubits+1)

    for q in range(qubits):
        cir.h(q)

    for q in range(qubits-1,-1,-1):
        cir.cu1(np.pi*the_phase*2**(qubits-q),q,qubits)

    for q in range(qubits):
        cir.h(q)
        for tar in range(q+1,qubits):
            cir.cu1(-np.pi/(2**(tar-q)),q,tar)

    return cir.qasm()
\end{lstlisting}

\subsubsection{Grover's Algorithm}

\textbf{Description}. Grover's algorithm \cite{grover1996fast} is one of the most commonly used algorithms in quantum computing to search the solution of an integer function.

Suppose you are given an oracle $O$ such that $O\ket{x}=(-1)^{f(x)}\ket{x}$, then initialise the state to be the equal superposition state $\ket{\psi}=\frac{1}{N^{1/2}}\sum_{x=0}^{N-1}{\ket{x}}$, and iteratively apply the Grover operator 
$$G=(2\ket{\psi}\bra{\psi}-I)O,$$ and finally measure all the qubits, you will get a solution of the function $f(x)$ with a probability close to 1.

\begin{figure}[h]
\centering
\centerline{
\Qcircuit @C=0.7em @R=0.4em {
&\gate{H} &\ctrl{1} &\gate{H} &\gate{X} &\qw      &\ctrl{1} &\qw      &\gate{X} &\gate{H} &\qw\\
&\gate{H} &\ctrl{1} &\gate{H} &\gate{X} &\gate{H} &\targ    &\gate{H} &\gate{X} &\gate{H} &\qw\\
&\gate{H} &\targ    &\qw      &\qw      &\qw      &\qw      &\qw      &\qw      &\gate{H} &\qw\\
}
}
\caption{An example of Grover's algorithm. Here, $f(x)$ is a 2-bit function and $f(x)=x_1 \cdot x_2$.}
\label{cir_grover_alg}
\end{figure}
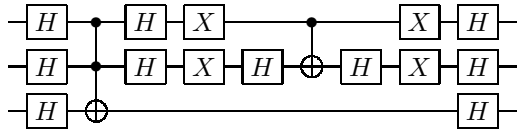

Figure \ref{cir_grover_alg} gives an example of the Grover's algorithm. Here, $f(x)=x_1 \cdot x_2$, running this circuit and measuring at the end will obtain $11$, which is the solution of this function.

{\vskip 3pt}

\noindent\textbf{OpenQASM Code}. The OpenQASM 2.0 description of this circuit is as follows.

\begin{lstlisting}[]
OPENQASM 2.0;
include "qelib1.inc";
qreg q[3];
creg c[3];
h q[0];
h q[1];
h q[2];
ccx q[0],q[1],q[2];
h q[0];
h q[1];
x q[0];
x q[1];
h q[1];
cx q[0],q[1];
h q[1];
x q[0];
x q[1];
h q[0];
h q[1];
h q[2];
\end{lstlisting}

\noindent\textbf{Generation Script}. The python code for generating circuits for arbitrary qubits Grover's search is as follows. In this code, you can assign the number of qubits $n$ of the search space, where the solution is $\ket{1}^{\otimes n}$. Then there will be one qubit serving as oracle workspace and $n-2$ qubits as ancilla qubits. Thus, the circuit will totally have $2n-1$ qubits and the $C^n(X)$ will be decomposed to a series of $CCX$ (\textit{i.e.} Toffoli) gates using the ancilla qubits.

\begin{lstlisting}[language = Python]
def gen_grover(qubits, r):
    cir = QuantumCircuit(2 * qubits - 1)
    # add H
    for q in range(qubits):
        cir.h(q)
    cir.h(2 * qubits - 2)
    for k in range(r):
        # add tofolli
        cir.ccx(0, 1, qubits)
        for q in range(2, qubits):
            cir.ccx(q, q + qubits - 2, q + qubits - 1)
        for q in range(qubits - 2, 1, -1):
            cir.ccx(q, q + qubits - 2, q + qubits - 1)
        if qubits > 2:
            cir.ccx(0, 1, qubits)
            # add H
        for q in range(qubits):
            cir.h(q)
        # add X
        for q in range(qubits):
            cir.x(q)
        cir.h(qubits - 1)
        if qubits == 2:
            cir.cx(0, 1)
        elif qubits == 3:
            cir.ccx(0, 1, 2)
        else:
            cir.ccx(0, 1, qubits)
            for q in range(2, qubits - 2):
                cir.ccx(q, q + qubits - 2, q + qubits - 1)
            cir.ccx(qubits - 2, 2 * qubits - 4, qubits - 1)

            for q in range(qubits - 3, 1, -1):
                cir.ccx(q, q + qubits - 2, q + qubits - 1)
            if qubits > 2:
                cir.ccx(0, 1, qubits)
        cir.h(qubits - 1)
        # add X
        for q in range(qubits):
            cir.x(q)
            # add H
        for q in range(qubits):
            cir.h(q)

    cir.h(2 * qubits - 2)
    return cir.qasm()
\end{lstlisting}

\subsubsection{Quantum Adder}

\textbf{Description}. Quantum adder is a quantum circuit to implement add operation on two bit strings. For example, if we compute '2+3=5', then we represent the input string as '010' and '011', and the expected output bit string is '101'. The implementation of the quantum adder circuit is illustrated as follows, and the additional qubits are used to store the carry bit~\cite{4786689}. 

{\vskip 3pt}

\noindent\textbf{OpenQASM Code}. The OpenQASM description of 3-bit adder circuit is shown as follows.

\begin{figure}[h]
\centering
\centerline{
\scalebox{1.0}{
\Qcircuit @C=1.0em @R=0.8em @!R { \\
	 	\nghost{ {q}_{0} :  } & \lstick{ {q}_{0} :  } & \qw & \qw & \ctrl{2} & \ctrl{2} & \qw & \qw & \qw & \qw\\ 
	 	\nghost{ {q}_{1} :  } & \lstick{ {q}_{1} :  } & \ctrl{1} & \ctrl{1} & \qw & \qw & \qw & \qw & \qw & \qw\\ 
	 	\nghost{ {q}_{2} :  } & \lstick{ {q}_{2} :  } & \ctrl{1} & \targ & \ctrl{1} & \targ & \qw & \qw & \qw & \qw\\ 
	 	\nghost{ {q}_{3} :  } & \lstick{ {q}_{3} :  } & \targ & \qw & \targ & \ctrl{2} & \ctrl{2} & \qw & \qw & \qw\\ 
	 	\nghost{ {q}_{4} :  } & \lstick{ {q}_{4} :  } & \ctrl{1} & \ctrl{1} & \qw & \qw & \qw & \qw & \qw & \qw\\ 
	 	\nghost{ {q}_{5} :  } & \lstick{ {q}_{5} :  } & \ctrl{1} & \targ & \qw & \ctrl{1} & \targ & \qw & \qw & \qw\\ 
	 	\nghost{ {q}_{6} :  } & \lstick{ {q}_{6} :  } & \targ & \qw & \qw & \targ & \ctrl{2} & \ctrl{2} & \qw & \qw\\ 
	 	\nghost{ {q}_{7} :  } & \lstick{ {q}_{7} :  } & \ctrl{1} & \ctrl{1} & \qw & \qw & \qw & \qw & \qw & \qw\\ 
	 	\nghost{ {q}_{8} :  } & \lstick{ {q}_{8} :  } & \ctrl{1} & \targ & \qw & \qw & \ctrl{1} & \targ & \qw & \qw\\ 
	 	\nghost{ {q}_{9} :  } & \lstick{ {q}_{9} :  } & \targ & \qw & \qw & \qw & \targ & \qw & \qw & \qw\\ 
\\ }}
}
\caption{Circuit of 3-bit adder}
\end{figure}
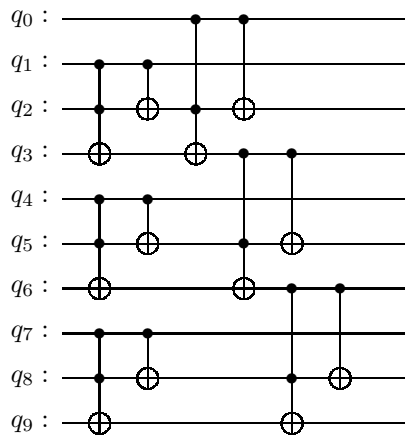

\begin{lstlisting}[]
OPENQASM 2.0;
include "qelib1.inc";
qreg q[10];
ccx q[1],q[2],q[3];
cx q[1],q[2];
ccx q[4],q[5],q[6];
cx q[4],q[5];
ccx q[7],q[8],q[9];
cx q[7],q[8];
ccx q[0],q[2],q[3];
ccx q[3],q[5],q[6];
ccx q[6],q[8],q[9];
cx q[0],q[2];
cx q[3],q[5];
cx q[6],q[8];
\end{lstlisting}

\noindent\textbf{Generation Script}. The python code for generating an adder circuit is as follows, the function takes the number of qubits as the input and outputs the OpenQASM text of the circuit. 

\begin{lstlisting}[language = Python]
def header(nq):
    str = ""
    str += "OPENQASM 2.0;\n"
    str += "include \"qelib1.inc\";\n"
    str += "qreg q[%d];\n"%(nq)
    return str

def carry_gate(s):
    str=''
    str+="ccx q[%d],q[%d],q[%d];\n"%(s+1, s+2, s+3)
    str+="cx q[%d],q[%d];\n"%(s+1, s+2)
    return str

def adder(numq):
    str = "//Adder with %d qubits input.\n"%numq
    str += header(3*numq+1)
    for i in range(numq):
        str+=carry_gate(3*i)
    for i in range(numq):
        str+="ccx q[%d],q[%d],q[%d];\n"%(3*i, 3*i+2, 3*i+3)
    for i in range(numq):
        str+="cx q[%d],q[%d];\n"%(3*i, 3*i+2)
    return str
\end{lstlisting}

\subsection{Reversible Circuits}\label{sec-reversible}
\textbf{Description.} A classical \(n\)-bit reversible gate is a bijective mapping \(f\) from the set \(\{0,1\}^n\) of \(n\)-bit data onto itself. Thus the vector of input states can always be reconstructed from the vector of output states. A combinational logic circuit is reversible if it only contains reversible gates and has no fan-out. Classical reversible circuits may be implemented in quantum technology and have important applications in many quantum algorithms such as the arithmetic module of Shor's Algorithm and the oracle of Grover's Algorithm.

We collect the classical reversible circuits in the Reversible Logic Synthesis Benchmarks Page~\cite{RLSBP}. The following are the elementary gates they used:

\begin{definition}
A generalized Toffoli gate TOF(\(x_1\), \(x_2\), ..., \(x_n\); \(x_{n+1}\)) is a gate which maps a Boolean pattern (\(x_1\), \(x_2\), ..., \(x_n\), \(x_{n+1}\)) to (\(x_1\), \(x_2\), ..., \(x_n\), \(x_{n+1}+x_1x_2\ldots x_n\)), where "+" is a modula-2 addition.
\end{definition}
Examples:
\begin{enumerate}
\item NOT gate is a TOF(\(\emptyset;a\)) gate.
\item CNOT gate is a TOF(\(a;b\)) gate.
\item Original Toffoli gate is a TOF(\(a,b;c\)).
\end{enumerate}

\begin{definition}
A generalized Fredkin gate FRE(\(x_1\), \(x_2\), ..., \(x_n\); \(x_{n+1}\), \(x_{n+2}\)) is a gate which maps Boolean pattern (\(x_1\), \(x_2\), ..., \(x_n\), \(x_{n+1}\), \(x_{n+2}\)) to (\(x_1\), \(x_2\), ..., \(x_n\), \(x_{n+2}\), \(x_{n+1}\)) if and only if Boolean product \(x_1x_2\ldots x_n = 1\), otherwise the pattern is unchanged.
\end{definition}
Examples:
\begin{enumerate}
\item SWAP gate is a FRE(\(\emptyset;a,b\)) gate.
\item Original Fredkin gate is a FRE(\(a;b,c\)) gate.
\end{enumerate}


\noindent\textbf{OpenQASM Code.} We translate their circuit description into the OpenQASM 2.0 format. The following gives an example of the reversible 5-bit adder circuit.
\begin{lstlisting}[]
OPENQASM 2.0;
include "qelib1.inc";
qreg q[11];
cx q[3], q[2];
cx q[5], q[4];
cx q[7], q[6];
cx q[9], q[8];
cx q[9], q[10];
cx q[7], q[9];
cx q[5], q[7];
cx q[3], q[5];
ccx q[0], q[1], q[3];
ccx q[2], q[3], q[5];
ccx q[4], q[5], q[7];
ccx q[6], q[7], q[9];
ccx q[8], q[9], q[10];
cx q[9], q[8];
ccx q[6], q[7], q[9];
cx q[7], q[6];
ccx q[4], q[5], q[7];
cx q[5], q[4];
ccx q[2], q[3], q[5];
cx q[3], q[2];
ccx q[0], q[1], q[3];
cx q[3], q[5];
cx q[5], q[7];
cx q[7], q[9];
cx q[1], q[0];
cx q[3], q[2];
cx q[5], q[4];
cx q[7], q[6];
cx q[9], q[8];
\end{lstlisting}

\subsection{Qubit Mapping}\label{sec-qubitmapping}

\textbf{Description. }On the current superconducting quantum processors, 2-qubit gates are usually  unavailable for arbitrary pairs of qubits but only for a small part of them. In order to make all the 2-qubit gates in a circuit available on a specific quantum chip, we have to map the qubits in the circuit to those on the quantum chip and insert some SWAP gates. Meanwhile, we want to make sure that the modified circuit is optimal on depth or the number of inserted SWAP gates. However, since the above problem, known as the qubit mapping problem, is NP-complete, it is difficult to theoretically evaluate the performance of different algorithms. Instead, the performance can be evaluated through benchmarks. \cite{tan2020optimality} presents an algorithm to generate benchmarks of the qubit mapping problems on specific quantum processors along with optimal solutions. We have implemented the generating algorithm in Python. The input and output of the implementation are described below:

The set of qubit pairs which are available for 2-qubit gates is given by the edge set $E$ of a graph $G$. Given depth $d$, gate count $N$ and the proportion of 2-qubit gates $p_2$, a random quantum circuit which can be executed on the graph $G$ is generated along with an optimal qubit mapping of depth $d$. 

{\vskip 3pt}

\noindent\textbf{Generation Script}. An example of our implementation of QUEKO is shown below:

\begin{lstlisting}[language=Python]
from QUEKO import QUEKO

edges = [(0, 1), (1, 2), (1, 3), (3, 4)]
prob = QUEKO(edges=edges, depth=5, gateCount=10, p2=0.3)

# print the generated OpenQASM program
print(prob.qasm2) 

# print the optimal solution, where the $i$-th qubit 
# should be mapped to the $i$-th element of the list.
print("//" + prob.optimalMapping) 
\end{lstlisting}

{\vskip 3pt}

\noindent\textbf{OpenQASM Code}. A sample output of the above program is:

\begin{lstlisting}
OPENQASM 2.0;
include "qelib1.inc";
qreg q[5];
cx q[1], q[2];
t q[4];
x q[1];
h q[0];
z q[3];
cx q[1], q[3];
y q[3];
cx q[4], q[2];
h q[3];
y q[2];

//[2, 3, 1, 4, 0]
\end{lstlisting}

Note that the gate \lstinline{cx q[4], q[2]} is not available on the graph given by \lstinline{edges}. However, if we apply the map

$$
0 \mapsto 2,\ 1 \mapsto 3,\ 2\mapsto 1,\ 3\mapsto 4,\ 4\mapsto 0
$$

to the qubits in the above program, all the $cx$ are then available on the graph.

\subsection{Random Circuits}
\subsubsection{Random Clifford Circuits}
\textbf{Description}. Clifford operation plays an important role in quantum error correction, randomized benchmarking protocols and quantum circuit simulation. By definition, Clifford operation is a unitary operation taking elements of \(G_n\) to elements of \(G_n\), where \(G_n\) is the Pauli group on \(n\) qubits. Any \(n\)-qubit Clifford operation can be simulated using \(O(n^2)\) Hadamard, phase and controlled-NOT gates. Clifford group elements are important and frequently encountered subsets of physical-level and fault-tolerant quantum circuits~\cite{bravyi2021clifford}, and sometimes an entire quantum algorithm can be a Clifford circuit (e.g., Bernstein–Vazirani~\cite{nielsen2002quantum}).

The Clifford group is a unitary 2-design. That is, a random uniformly distributed element of the Clifford group has exactly the same second order moments as the Haar random unitary operation. This means the random Clifford operations can serve as a substitute for Haar random unitaries in any application that depends only on the second order moments. In Qiskit, the random Clifford is sampled using the method of \cite{bravyi2021hadamard}. Then the Clifford circuit is synthesized by the method in \cite{aaronson2004improved} and optimized by the method in \cite{bravyi2021clifford}.

{\vskip 3pt}

\noindent\textbf{Generation Script}. The script for generating random Clifford circuit is as follows.

\begin{lstlisting}[language=Python]
from qiskit import quantum_info

def gen_rand_cliff(n):

    cliff=quantum_info.random_clifford(n)
    AG=quantum_info.decompose_clifford(cliff,method='AG')
    GD=quantum_info.decompose_clifford(cliff,method='greedy')
    
    return AG.qasm(),GD.qasm()
\end{lstlisting}

{\vskip 3pt}

\noindent\textbf{OpenQASM Code}. The following gives a 3-qubit random Clifford circuit in OpenQASM 2.0 format.

\begin{lstlisting}[]
OPENQASM 2.0;
include "qelib1.inc";
qreg q[3];
s q[0];
h q[0];
s q[1];
h q[1];
s q[1];
h q[2];
cx q[2],q[1];
cx q[1],q[0];
cx q[0],q[2];
h q[1];
s q[2];
h q[2];
cx q[2],q[1];
s q[2];
x q[0];
z q[1];
\end{lstlisting}

\subsubsection{Quantum Volume}

\textbf{Description}. Quantum volume \cite{cross2019validating} is a metric that can be used to measure the capabilities and error rates of a quantum computer. It quantifies the largest random circuit of equal width and depth that the computer can successfully implement. The circuit model used for measuring quantum volume is as follows.

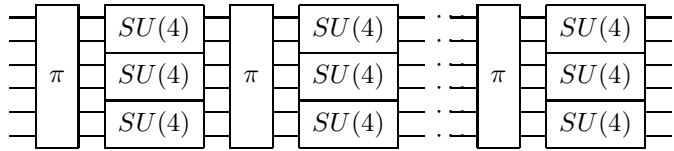
\begin{figure}[h]
\centering
\centerline{
\Qcircuit @C=1em @R=0em {
& \multigate{5}{\pi} & \multigate{1}{SU(4)} & \multigate{5}{\pi} & \multigate{1}{SU(4)} &\qw &\cdots & \multigate{5}{\pi} & \multigate{1}{SU(4)} &\qw\\
& \ghost{\pi}        & \ghost{SU(4)}        & \ghost{\pi}        & \ghost{SU(4)}        &\qw &\cdots & \ghost{\pi}        & \ghost{SU(4)}        &\qw\\
& \ghost{\pi}        & \multigate{1}{SU(4)} & \ghost{\pi}        & \multigate{1}{SU(4)} &\qw &\cdots & \ghost{\pi}        & \multigate{1}{SU(4)} &\qw\\
& \ghost{\pi}        & \ghost{SU(4)}        & \ghost{\pi}        & \ghost{SU(4)}        &\qw &\cdots & \ghost{\pi}        & \ghost{SU(4)}        &\qw\\
& \ghost{\pi}        & \multigate{1}{SU(4)} & \ghost{\pi}        & \multigate{1}{SU(4)} &\qw &\cdots & \ghost{\pi}        & \multigate{1}{SU(4)} &\qw\\
& \ghost{\pi}        & \ghost{SU(4)}        & \ghost{\pi}        & \ghost{SU(4)}        &\qw &\cdots & \ghost{\pi}        & \ghost{SU(4)}        &\qw\\
}
}
\caption{The circuit model for quantum volume.}
\label{cir_qv_alg}
\end{figure}

Here, $\pi$ is a permutation of qubits, and every $SU(4)$ represents a 2-qubit unitary gate sampled from the Haar measure on SU(4). There will be $d$ layers of this module if the depth of the circuit is $d$, and if the number of qubits in the circuit is odd, then every layer will have an idle qubit. 

In our benchmark, we give a series of circuits for quantum volume constructed on basic quantum gates. And the $SU(4)$ gates are decomposed using qiskit.

{\vskip 3pt}

\noindent\textbf{OpenQASM Code}. The following is an example of the quantum volume circuit with qubits 2 depth 2, and these circuits are decomposed to $cx$ and $u3$ gate using the standard QASM representation.

\begin{lstlisting}[]
OPENQASM 2.0;
include "qelib1.inc";
qreg q[2];
creg c[2];
u3(0.10690831,-1.8026866,0.78825838) q[0];
u3(2.4950187,-3.5814199,2.5427764) q[1];
cx q[0],q[1];
u3(0.82807016,-3*pi/2,pi/2) q[0];
u3(pi/2,-pi,-2.8687452) q[1];
cx q[0],q[1];
u3(0.20800243,0,-3*pi/2) q[0];
u3(pi/2,0,-3*pi/2) q[1];
cx q[0],q[1];
u3(2.0104954,1.3664259,-1.5218768) q[0];
u3(0.94734104,-0.84662394,3.2456914) q[1];
u3(1.143506,-3.1351223,-0.55632777) q[0];
u3(1.3587079,-3.1777958,0.27175853) q[1];
cx q[0],q[1];
u3(1.0096868,-3*pi/2,pi/2) q[0];
u3(pi/2,-pi,-3.0848291) q[1];
cx q[0],q[1];
u3(0.61517883,0,-3*pi/2) q[0];
u3(pi/2,0,-3*pi/2) q[1];
cx q[0],q[1];
u3(0.09341401,-1.4199754,4.7299314) q[0];
u3(0.45442386,-2.4437295,1.2711765) q[1];
\end{lstlisting}

\noindent\textbf{Generation Script}. The script for generating circuit for arbitrary qubits and arbitrary depth is as follows.

\begin{lstlisting}[language = Python]
from qiskit.quantum_info.synthesis \
import two_qubit_cnot_decompose
def random_SU(n):
    X = (np.random.randn(n, n) + 1j * np.random.randn(n, n))
    Q, R = linalg.qr(X)
    Q /= pow(linalg.det(Q), 1/n)
    return Q

def gen_qv(qubits, depth):
    cir = QuantumCircuit(qubits, qubits)
    for j in range(depth):
        perm = np.random.permutation(qubits)
        for k in range(qubits // 2):
            q = [int(perm[2 * k]), int(perm[2 * k + 1])]
            SU = random_SU(4)
            decomposed_SU = two_qubit_cnot_decompose(SU)
            for gate in decomposed_SU:
                i0 = q[gate[1][0].index]
                if gate[0].name == "cx":
                    i1 = q[gate[1][1].index]
                    qv.cx(i0, i1)
                elif gate[0].name == "u1":
                    qv.u1(gate[0].params[2], i0)
                elif gate[0].name == "u2":
                    qv.u2(gate[0].params[1], \
                          gate[0].params[2], i0)
                elif gate[0].name == "u3":
                    qv.u3(gate[0].params[0], \
                    gate[0].params[1],gate[0].params[2], i0)
                elif gate[0].name == "id":
                    pass
    return cir.qasm()
\end{lstlisting}

\subsubsection{Supremacy Circuits}

Random circuits have been widely used in works related to the quantum supremacy \cite{boixo2018characterizing}. Since there are already some benchmarks with such circuits, we just include the existing and commonly used circuits directly in our benchmark. One of such benchmark is GRCS \footnote{\url{https://github.com/sboixo/GRCS}}. GRCS provides a lot of random circuits, but all in a '.txt' format and not given in the standard OpenQASM language. In our benchmark, we give the OpenQASM version of the random circuits shown in GRCS.

\section{Dynamic Quantum Circuits}\label{sec:dynamic}
Dynamic quantum circuit is a model of quantum computation, in which quantum algorithms can be executed in a more flexible classical-quantum hybrid way, which can sometimes reduce the costs of quantum resources. The dynamic quantum circuit is adopted as an alternative and beneficial way for executing quantum algorithms on NISQ devices. Our benchmark includes a series of dynamic quantum circuits.

The biggest difference between dynamic quantum circuits and conventional quantum circuits is that measurements will usually appear in the middle of the circuit, and a series of classically controlled gates will be applied according to the results of the measurements.

\subsection{Quantum Teleportation}\label{teleportaion}
\textbf{Description}. One of the simplest examples of dynamic quantum circuits is teleportation. Teleportation is a protocol for transmitting a qubit between two users by sending two classical bits of information \cite{bennett1993teleporting}. 
{\vskip 3pt}

\noindent\textbf{OpenQASM Code}.  The corresponding circuit and the OpenQASM 2.0 description are as follows. The first qubit belongs to Alice and the last qubit belongs to Bob. By first applying a series of quantum gates and sending the measurement result to Bob, Bob can obtain the state of the first qubit of Alice by applying a series of gates according to the measurement information.

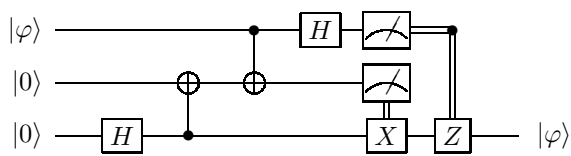
\begin{figure}[h]
\centerline{
\Qcircuit @C=.7em @R=.2em @! {
\lstick{\ket{\varphi}} & \qw      & \qw       & \ctrl{1} &\gate{H} & \meter         & \control \cw\\
\lstick{\ket{0}}       & \qw      & \targ     & \targ    & \qw     & \meter         & \cwx         \\
\lstick{\ket{0}}       & \gate{H} & \ctrl{-1} & \qw      & \qw     & \gate{X} \cwx  & \gate{Z} \cwx &\qw &\lstick{\ket{\varphi}}
}
}
\caption{Dynamic quantum circuit for Teleportation.}
\label{cir_dqc_tele}
\end{figure}

\begin{lstlisting}[]
OPENQASM 2.0;
include "qelib1.inc";
qreg q[3];
creg c0[1];
creg c1[1];
creg c2[1];
h q[2];
cx q[2],q[1];
cx q[0],q[1];
h q[0];
measure q[1] -> c1[0];
if(c1==1) x q[2];
measure q[0] -> c0[0];
if(c0==1) z q[2];
\end{lstlisting}

\subsection{Semiclassical Fourier Transform}
\begin{figure}[h]
\centering
\centerline{
\Qcircuit @C=0.7em @R=0.4em {
\lstick{\ket{j_1}}  & \gate{H} &\meter\cwx[1] &\cw      &\control \cw\\
\lstick{\ket{j_2}}  & \qw      &\gate{R_2}    &\gate{H} &\qw \cwx  &\meter\cwx[1]\\
\lstick{\ket{j_3}}  & \qw      &\qw           &\qw      &\gate{R_3}\cwx      &\gate{R_2}    &\gate{H} &\meter\\
}
}
\caption{Dynamic quantum circuit for Fourier transform.}
\label{cir_dqc_qft_alg}
\end{figure}
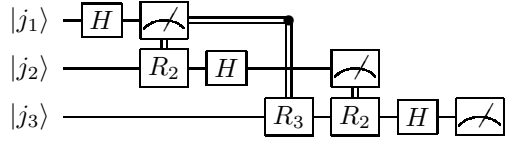

\textbf{Description}. The quantum Fourier transform can also be represented as a dynamic quantum circuit form \cite{griffiths1996semiclassical}. The following gives the detailed information of the dynamic version of the quantum Fourier transform. From Figure \ref{cir_qft_alg} and Figure \ref{cir_dqc_qft_alg} you can see that all the controlled-$R_k$ gates are replaced by a measurement and classically controlled $R_k$ gates.

The advantage of using this dynamic quantum circuit is that there is no need to apply 2-qubit gates during the process, all the 2-qubit gates can be replaced by single-qubit gates chosen according to the measurement results.

{\vskip 3pt}

\noindent\textbf{OpenQASM Code}. Following gives the OpenQASM description of the dynamic version of quantum Fourier transform.
\begin{lstlisting}[]
OPENQASM 2.0;
include "qelib1.inc";
qreg q[3];
creg c0[1];
creg c1[1];
creg c2[1];
h q[0];
measure q[0] -> c0[0];
if(c0==1) u1(pi/2) q[1];
if(c0==1) u1(pi/4) q[2];
h q[1];
measure q[1] -> c1[0];
if(c1==1) u1(pi/2) q[2];
h q[2];
measure q[2] -> c2[0];
\end{lstlisting}

\noindent\textbf{Generation Script}. Similar to the conventional quantum circuit, the corresponding circuit can be generated as follows.

\begin{lstlisting}[language = Python]
def gen_dqc_qft(qubits):

    cir= QuantumCircuit(qubits)

    for q in range(qubits):
        cir.h(q)
        c= ClassicalRegister(1,'c'+str(q))
        cir.add_register(c)
        cir.measure(q,c)
        for tar in range(q + 1, qubits):
            theta = np.pi / 2 ** (tar - q)
            cir.u1(theta,tar).c_if(c, 1)

    return cir.qasm()
\end{lstlisting}

\subsection{Iterative Phase Estimation}
\textbf{Description}. Since the quantum phase estimation uses the inverse quantum Fourier transform as a subroutine, it can also be executed as a dynamic quantum circuit. The dynamic version of quantum phase estimation is as follows.

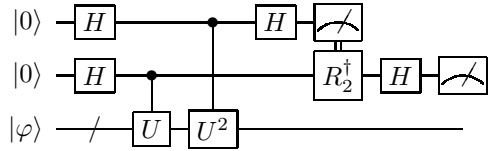
\begin{figure}[h]
\centering
\centerline{
\Qcircuit @C=0.7em @R=0.4em {
\lstick{\ket{0}}  & \gate{H} &\qw      &\ctrl{2}  &\gate{H} &\meter\\
\lstick{\ket{0}}  & \gate{H} &\ctrl{1} &\qw       &\qw      &\gate{R_2^{\dagger}} \cwx &\gate{H}&\meter\\
 \lstick{\ket{\varphi}} & {/}\qw      &\gate{U} &\gate{U^2}&\qw      &\qw&\qw&\qw\\
}
}
\caption{Quantum circuit for phase estimation.}
\label{cir_dqc_qpe_alg}
\end{figure}

The advantage of the phase estimation implemented by the dynamic quantum circuit is that the circuit can be executed with only 2-qubit gates, and more details are shown in \cite{corcoles2021exploiting}.

{\vskip 3pt}

\noindent\textbf{OpenQASM Code}.  The following gives the dynamic version of 3-qubit phase estimation.

\begin{lstlisting}[]
OPENQASM 2.0;
include "qelib1.inc";
qreg q[3];
creg c0[1];
creg c1[1];
h q[0];
h q[1];
cu1(pi/512) q[1],q[2];
cu1(pi/256) q[0],q[2];
h q[0];
measure q[0] -> c0[0];
if(c0==1) u1(-pi/2) q[1];
h q[1];
measure q[1] -> c1[0];
\end{lstlisting}

\noindent\textbf{Generation Script}.  The code for generating dynamic quantum phase estimation circuit of any qubit number is as follows.

\begin{lstlisting}[language = Python]
def gen_dqc_pe(qubits, the_phase):
    cir = QuantumCircuit(qubits + 1)

    for q in range(qubits):
        cir.h(q)

    for q in range(qubits - 1, -1, -1):
        cir.cu1(np.pi*the_phase*2**(qubits-q),q,qubits)

    for q in range(qubits):
        cir.h(q)
        c = ClassicalRegister(1, 'c' + str(q))
        cir.add_register(c)
        cir.measure(q, c)
        for tar in range(q + 1, qubits):
            theta = np.pi / 2 ** (tar - q)
            cir.u1(-theta, tar).c_if(c, 1)

    return cir.qasm()
\end{lstlisting}

\subsection{Quantum Error correction}
\textbf{Description}. Another example of dynamic quantum circuits is the circuit for quantum error correction, where a quantum state must first experience a period of syndrome measurement, and then be recovered according to the measurement result. 

{\vskip 3pt}

\noindent\textbf{OpenQASM Code}.  The following gives the circuit model of error correction as well as the OpenQASM 2.0 description of the bit-flip and phase-flip code \cite{nielsen2002quantum}.

\begin{figure}[h]
\centerline{
\Qcircuit @C=1.3em @R=.6em {
& \qw & \qw & \ctrl{3} & \qw & \qw & \qw &
\ctrl{5} & \qw & \qw &
\multigate{2}{\ \mathcal{R}\ } & \qw\\
& \qw & \qw & \qw & \ctrl{2} & \ctrl{3} & \qw &
\qw & \qw & \qw & \ghost{\ \mathcal{R}\ } \qw &
\qw\\
& \qw & \qw & \qw & \qw & \qw & \ctrl{2} & \qw &
\ctrl{3} & \qw & \ghost{\ \mathcal{R}\ } \qw &
\qw\\
& & \lstick{\ket{0}} & \targ \qw & \targ \qw &
\qw & \qw & \qw & \qw & \meter &
\control \cw \cwx\\
& & \lstick{\ket{0}} & \qw & \qw & \targ \qw &
\targ \qw & \qw & \qw & \meter &
\control \cw \cwx\\
& & \lstick{\ket{0}} & \qw & \qw & \qw & \qw &
\targ \qw & \targ \qw & \meter
&\control \cw \cwx
}
}
\caption{A circuit model for Quantum Error correction.}
\label{cir_dqc_error_correc}
\end{figure}
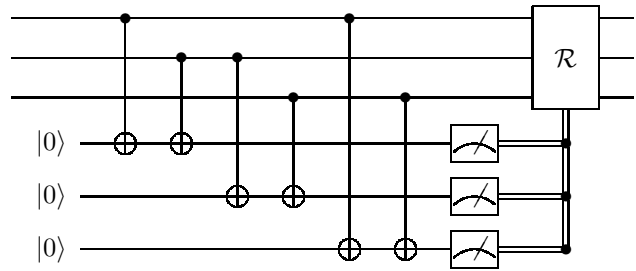

\begin{lstlisting}[]
OPENQASM 2.0;
include "qelib1.inc";
qreg q[6];
creg c[6];
cx q[0],q[3];
cx q[1],q[3];
cx q[1],q[4];
cx q[2],q[4];
cx q[0],q[5];
cx q[2],q[5];
measure q[3] -> c[3];
measure q[4] -> c[4];
measure q[5] -> c[5];
if(c==5) x q[0];
if(c==6) x q[1];
if(c==3) x q[2];
\end{lstlisting}

\begin{lstlisting}[]
OPENQASM 2.0;
include "qelib1.inc";
qreg q[6];
creg c[6];
h q[0];
h q[1];
h q[2];
cx q[0],q[3];
cx q[1],q[3];
cx q[1],q[4];
cx q[2],q[4];
cx q[0],q[5];
cx q[2],q[5];
measure q[3] -> c[3];
measure q[4] -> c[4];
measure q[5] -> c[5];
if(c==5) x q[0];
if(c==6) x q[1];
if(c==3) x q[2];
h q[0];
h q[1];
h q[2];
\end{lstlisting}

\subsection{State Injection}
\textbf{Description}. State injection is the technique for implementing some quantum gates using a dynamic scheme. The basic idea is that the effect of this gate can be implemented by using a special state and a series of gates that can be implemented more simply.

{\vskip 3pt}

\noindent\textbf{OpenQASM Code}.  The following gives the circuit and the OpenQASM scripts for the dynamic implementation of the $S$ and $T$ gate \cite{ryan2017hardware}.

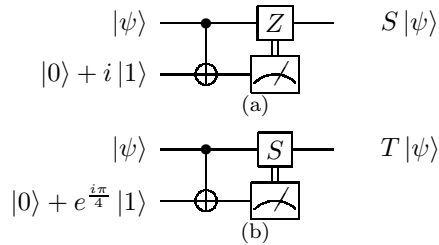
\begin{figure}[h]
\subfigure[]{
\centerline{
\Qcircuit @C=1.3em @R=.6em {
\lstick{\ket{\psi}}       & \ctrl{1} &\gate{Z}  &\qw &\rstick{S\ket{\psi}}\\
\lstick{\ket{0}+i\ket{1}} & \targ    &\meter \cwx
}
}
}
\subfigure[]{
\centerline{
\Qcircuit @C=1.3em @R=.6em {
\lstick{\ket{\psi}}       & \ctrl{1} &\gate{S}  &\qw &\rstick{T\ket{\psi}}\\
\lstick{\ket{0}+e^{\frac{i\pi}{4}}\ket{1}} & \targ    &\meter \cwx
}
}
}
\caption{The quantum circuits for state injection, (a) implementation of $S$ gate, (b) implementation of $T$ gate.}
\label{cir_dqc_state_injection}
\end{figure}

\begin{lstlisting}[]
OPENQASM 2.0;
include "qelib1.inc";
qreg q[2];
creg c0[1];
creg c1[1];
h q[0];
h q[0];
cx q[0],q[1];
measure q[1] -> c1[0];
if(c1==1) z q[0];
\end{lstlisting}

\begin{lstlisting}[]
OPENQASM 2.0;
include "qelib1.inc";
qreg q[2];
creg c0[1];
creg c1[1];
h q[0];
h q[0];
cx q[0],q[1];
measure q[1] -> c1[0];
if(c1==1) u1(pi/2) q[0];
\end{lstlisting}

\section{Sequential Quantum Circuits}\label{sec:sequential}

Sequential quantum circuits \cite{LP09,WLY21} are a new breed of quantum circuits that incorporate a clock signal. 
They can be understood as a generalization of classical (synchrounous) logic circuits. 
From another point of view, sequential quantum circuits are a special type of dynamic quantum circuits, since intermediate measurements are intended to perform at the end of each time step.

Sequential quantum circuits are useful in describing algorithms with loops and quantum feedback, e.g., Repeat-Until-Success circuits \cite{PS14} and quantum walks \cite{Kem03}. 
In this section, we select some of the examples of sequential quantum circuits and explain how we benchmark them in OpenQASM. 

\subsection{Repeat-Until-Success}

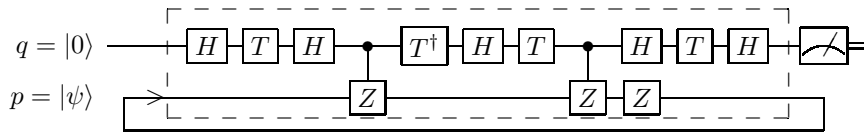
\begin{figure}[htp]\centering
    \normalsize\begin{equation*}\qquad \quad
    \Qcircuit @C=.6em @R=.6em {
    \lstick{q = \ket 0}    & \qw & \qw & \qw   & \qw & \gate{H} & \gate{T} & \gate{H} & \ctrl{1}     & \gate{T^\dag} & \gate{H} & \gate{T} & \ctrl{1}     & \gate{H} & \gate{T} & \gate{H} & \qw & \meter    & \cw  \\
    \lstick{p = \ket \psi} &     & \qw & \qw > & \qw & \qw      & \qw      & \qw      & \gate{Z}     & \qw           & \qw      & \qw      & \gate{Z}     & \gate{Z} & \qw      & \qw      & \qw & \qw \\
                  & \qwx  & \qw & \qw & \qw & \qw & \qw & \qw & \qw & \qw & \qw & \qw & \qw & \qw & \qw & \qw & \qw & \qw \qwx
                  \gategroup{1}{6}{2}{16}{1.5em}{--}
    }
    \end{equation*}
    \caption{A sequential quantum circuit for repeat-until-success implementation of $V_3$. This figure is taken from \cite{WLY21}.}
    \label{fig:rus}
    \end{figure}

    \textbf{Description}. Repeat-Until-Success circuits \cite{PS14} are a bunch of efficient implementations of quantum logic gates which make good use of intermediate measurements as feedback. 
    Figure \ref{fig:rus} shows a sequential quantum circuit for repeat-until-success implementation of quantum gate
    \[
    V_3 = \frac{I + 2iZ}{\sqrt{5}}.
    \]
    On this basis, we design several benchmarks.

    {\vskip 3pt}
    
    \noindent\textbf{OpenQASM Code}. In Figure \ref{fig:rus}, $q$ is the only input variable (qubit) while $p$ is the only internal variable (qubit). At each time step, the qubit $q$ will be measured in the computational basis indicating whether the implementation succeeds. We abstract the main process of the repeat-until-success implementation of $V_3$ in OpenQASM as follows. 
    
\begin{lstlisting}[]
def qrus() qubit:q, qubit:p
{
	h q;
	t q;
	h q;
	ctrl @ z q, p;
	inv @ t q;
	h q;
	t q;
	ctrl @ z q, p;
	h q;
	t q;
	h q;
	z p;
	bit result;
	result = measure q;
}
\end{lstlisting}
    
    Next, we will show how to use the function above to execute the sequential quantum circuit for one time step.
    Before the main process of the sequential circuit, we need to initialize the internal variable, which is qubit $p$ in our case. The following code sets $p$ to $\ket{+}$.
    
\begin{lstlisting}[]
qubit p;
h p; // p = |+>
\end{lstlisting}
    
    Every time before calling the function \texttt{qrus()}, we need to prepare input qubit $q$ beforehand. 
    
\begin{lstlisting}[]
qubit q;
reset q; // q = |0>
\end{lstlisting}
    
    Now we are ready to call \texttt{qrus()}.
    
\begin{lstlisting}[]
qrus(q, p);
\end{lstlisting}
    
    After calling \texttt{qrus(q, p)}, you can  obtain the measurement outcome of the sequential quantum circuit as follows.
    
\begin{lstlisting}[]
bit x;
x = measure q;
\end{lstlisting}
    
    Here, we note that qubit $q$ has already been measured in the execution of \texttt{qrus(p, q)} before that of \texttt{x = measure q;}. Nevertheless, the measurement outcome is always retrieved from a correct probability distribution. The reason we encapsulate measurements in the main process of the sequential circuit (i.e., \texttt{qrus()} in our case) is to ensure that a measurement must be performed at each time step. 

    \subsection{Quantum Walk}

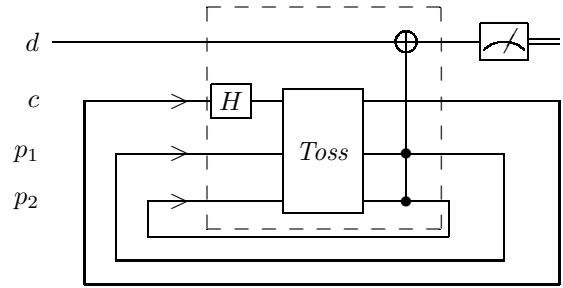
\begin{figure}[htp] \centering
\normalsize\begin{equation*}
\Qcircuit @C=1.2em @R=0.9em {
\lstick{d}            & \qw & \qw & \qw & \qw & \qw         & \qw               & \targ     & \qw & \meter    & \cw  \\
\lstick{c}        &     & \qw & \qw & \qw > & \gate{H}    & \multigate{2}{\mathit{Toss}}  & \qw       & \qw       & \qw & \qw  \\
\lstick{p_1}  & \qwx&     & \qw & \qw > & \qw         & \ghost{\mathit{Toss}}         & \ctrl{-2} & \qw       & \qw & \qwx\\
\lstick{p_2}  & \qwx& \qwx&     & \qw > & \qw         & \ghost{\mathit{Toss}}         & \ctrl{-3} & \qw       & \qwx& \qwx \\
                            & \qwx& \qwx& \qwx& \qw & \qw         & \qw               & \qw       & \qw \qwx  & \qwx&\qwx \\
                            & \qwx& \qwx& \qw & \qw & \qw         & \qw               & \qw       & \qw       & \qw \qwx&\qwx \\
                            & \qwx& \qw & \qw & \qw & \qw         & \qw               & \qw       & \qw       & \qw & \qw \qwx
                            \gategroup{1}{6}{4}{8}{1.8em}{--}
}
\end{equation*}
\caption{A sequential quantum circuit for quantum walk. This figure is taken from \cite{WLY21}.}\label{fig:qwalk}
\end{figure}

    \textbf{Description}. Quantum walks \cite{ADZ93,Kem03,ABN+01} are quantum generalizations of classical random walks, and have many applications in quantum algorithms. As an illustrative example (see Figure \ref{fig:qwalk}), we consider a quantum walk on a circle (with positions $\ket{0}_p, \ket{1}_p, \ket{2}_p, \ket{3}_p$) with an absorbing boundary $\ket{3}_p$. Here, $\mathit{Toss}$ is the toss operator
    \begin{align*}
        \mathit{Toss}: 
        & \ket{0}_c \ket{i}_p \to \ket{0}_c \ket{\left(i+1\right) \bmod 4}_p, \\
        & \ket{1}_c \ket{i}_p \to \ket{1}_c \ket{\left(i-1\right) \bmod 4}_p,
    \end{align*}
    which translates the position conditioned on the coin state.
    
    \textbf{OpenQASM Code}. In Figure \ref{fig:qwalk}, $d$ is the only input variable (qubit), and $c, p_1, p_2$ are the output variables (qubits). The OpenQASM code for the sequential quantum circuit in Figure \ref{fig:qwalk} is as follows.
    
\begin{lstlisting}[]
OPENQASM 3.0;

include "stdgates.inc";

gate coin q
{
	h q;
}

gate shift0 p1, p2
{
	ctrl @ x p2, p1;
	x p2;
}

gate shift1 p1, p2
{
	x p2;
	ctrl @ x p2, p1;
}

gate toss c, p1, p2
{
	x c;
	ctrl @ shift0 c, p1, p2;
	x c;
	ctrl @ shift1 c, p1, p2;
}

def qwalk() qubit:d, qubit:c, qubit:p1, qubit:p2
{
	coin c;
	toss c, p1, p2;
	ctrl @ ctrl @ x p1, p2, d;
	bit result;
	result = measure d;
}
\end{lstlisting}
    
    Similar to the use of the code in repeat-until-success circuits, after initializing qubit $d$, a call to $\texttt{qwalk(d, c, p1, p2)}$ will result in one step of quantum walk.

    \subsection{Classical Control}
    
\begin{figure}[htp]\centering
    \normalsize\begin{equation*}
    \Qcircuit @C=1.2em @R=.9em {
    \lstick{d = \ket{0}}            & \qw & \qw & \qw & \qw   & \qw       & \qw       & \qw       & \qw       & \targ     & \qw & \meter    & \cw  \\
    \lstick{q_1}                & \qw & \qw & \qw & \qw   & \ctrl{1}  & \ctrlo{1} & \ctrlo{1} & \ctrl{1}  & \qw       & \qw & \meter    & \cw  \\
    \lstick{q_2}                & \qw & \qw & \qw & \qw   & \ctrl{1}  & \ctrlo{1} & \ctrl{2}  & \ctrlo{3} & \qw       & \qw & \meter    & \cw  \\
    \lstick{p_1}                &     & \qw & \qw & \qw > & \ctrl{1}  & \gate{H}  & \qw       & \qw       & \ctrl{-3} & \qw       & \qw & \qw  \\
    \lstick{p_2}                & \qwx&     & \qw & \qw > & \ctrl{1}  & \qw       & \gate{H}  & \qw       & \qw       & \qw       & \qw & \qwx\\
    \lstick{p_3}                & \qwx& \qwx&     & \qw > & \targ     & \qw       & \qw       & \gate{H}  & \qw       & \qw       & \qwx& \qwx \\
                                & \qwx& \qwx& \qwx& \qw   & \qw       & \qw       & \qw       & \qw       & \qw       & \qw \qwx  & \qwx&\qwx \\
                                & \qwx& \qwx& \qw & \qw   & \qw       & \qw       & \qw       & \qw       & \qw       & \qw       & \qw \qwx&\qwx \\
                                & \qwx& \qw & \qw & \qw   & \qw       & \qw       & \qw       & \qw       & \qw       & \qw       & \qw & \qw \qwx
                                \gategroup{1}{6}{6}{10}{1.0em}{--}
    }
    \end{equation*}
    \caption{A sequential quantum circuit for classical control. This figure is taken from \cite{WLY21}.}\label{fig:classical-control}
    \end{figure}
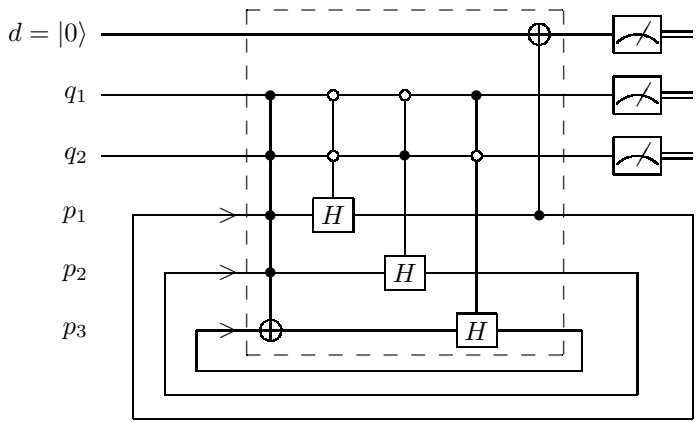

    \textbf{Description}. For the convenience to control the behaviour of quantum systems, we consider a class of sequential quantum circuits with their input variables in the computational basis. For example, in Figure \ref{fig:classical-control}, the input variables $q_1$ and $q_2$ control which kind of quantum gates is performed on $p_1, p_2, p_3$. In order to retrieve some information from the sequential quantum circuit, we use a detective qubit (an input variable) $d$ with it being initialized to $\ket{0}$.
    
    \textbf{OpenQASM Code}. The OpenQASM code for the sequential quantum circuit in Figure \ref{fig:classical-control} is as follows. 
    
\begin{lstlisting}[]
OPENQASM 3.0;

include "stdgates.inc";

def qctrl() qubit:d, qubit:q1, qubit:q2, qubit:p1, qubit:p2, qubit:p3
{
	ctrl(4) @ x q1, q2, p1, p2, p3;
	negctrl(2) @ q1, q2, p1;
	negctrl @ ctrl @ h q1, q2, p2;
	ctrl @ negctrl @ q1, q2, p3;
	cx p1, d;
	bit result[3];
	result[0] = measure d;
	result[1] = measure q1;
	result[2] = measure q2;
}
\end{lstlisting}
    
    After initializing input variables $d, q_1, q_2$, we can call \texttt{qctrl(d, q1, q2, d1, d2, d3)} for one time step of the computation.

\section{Variational Quantum Circuits}\label{sec:variational}
The benchmark on variational quantum circuits (VQCs, or parameterized quantum circuits) aims to provide VQC templates of variational quantum algorithms, including variational quantum eigensolver (VQE).

\subsection{Variational Quantum Eigensolver}
\textbf{Description}. Solving the ground state energy (minimum eigenvalue) of a Hamiltonian is fundamental in quantum chemistry and condensed matter physics.
By using various VQCs to model wavefunctions, VQE converts the minimum eigenvalue problem into optimization over parameters of VQCs. The reader can refer to a review~\cite{tilly2021variational} for comprehensive knowledge.

{\vskip 3pt}

\noindent\textbf{Generation Script}. In the benchmark, we provide a Python script named \tcode{BenchmarkVQE.py} to generate OpenQASM files for VQCs used in VQE literature.
The different types of VQCs (ansatz) we currently support and the corresponding arguments are as follows:

\begin{itemize}
    \item Hardware-efficient ansatz, \tcode{--ansatz hea}.
    \item Unitary coupled cluster (UCC) ansatz, \tcode{--ansatz ucc}.
    \item Symmetry-preserving ansatz, \tcode{--ansatz spa}.
\end{itemize}

After specifying the ansatz, there are some arguments (e.g., number of qubits) left to set. For detailed usage, please refer to our repository in GitHub.

For example, the following command
\begin{lstlisting}
python benchmark_vqe.py --ansatz hea --num_qubits 2
\end{lstlisting}
will generate an OpenQASM file \tcode{output.qasm} for hardware-efficient ansatz with $2$ qubits and all parameters randomly.

{\vskip 3pt}

\noindent\textbf{OpenQASM Code}. An example is as follows.
\begin{lstlisting}[]
OPENQASM 2.0;
include "qelib.inc";

qreg qs[2];

// Rotation
rz(2.3528045139656477) qs[0];
rx(5.056468340910666) qs[0];
rz(4.830262154311054) qs[0];
rz(5.826369114279527) qs[1];
rx(0.6640310016450327) qs[1];
rz(4.350196740814403) qs[1];
// Entangling
cx qs[0], qs[1];
// Rotation
rz(1.2300646901838261) qs[0];
rx(4.3476867872279685) qs[0];
rz(4.04916881465231) qs[0];
rz(3.1235589418914618) qs[1];
rx(3.5136963947870354) qs[1];
rz(2.656998011183734) qs[1];
// Entangling
cx qs[0], qs[1];
// Rotation
rz(1.7224607753970849) qs[0];
rx(5.561609408659551) qs[0];
rz(4.58779861660491) qs[0];
rz(6.2727331716861245) qs[1];
rx(4.711224351664875) qs[1];
rz(4.872654473631725) qs[1];
\end{lstlisting}

\section{Validation}\label{section_val}
All our benchmark circuits are validated by Qiskit~\cite{qiskit} and QCOR~\cite{qcor}. For OpenQASM 2.0 descriptions, Qiskit is used. To run an example named {\tt in\_file.qasm}, simply navigate to its directory and execute the codes in python:
\begin{lstlisting}
qc = qiskit.QuantumCircuit.from_qasm_file('./in_file.qasm')
simulator = qiskit.Aer.get_backend('aer_simulator')
qc = qiskit.transpile(qc,simulator)
result = simulator.run(qc, shots=1, memory=True).result()
\end{lstlisting}
For OpenQASM 3.0 descriptions, QCOR is used. To run an example named {\tt in\_file.qasm}, simply navigate to its directory and execute:
\begin{lstlisting}
qcor -shots 1024 in_file.qasm
./a.out
\end{lstlisting}
The simulation can be accelerated using TNQVM~\cite{tnqvm}, which leverages tensor network theory to simulate quantum circuits. To run the above example with TNQVM acceleration, simply execute:
\begin{lstlisting}
qcor -qpu tnqvm[tnqvm-visitor:exatn] -shots 1024 in_file.qasm
./a.out
\end{lstlisting}
For more details, we refer the readers to the official documentation and user guides~\cite{qcor-doc}.

\section{Example Applications}\label{section_app}
To demonstrate the utility, in this section, we discuss some  example applications and point out how the benckmark circuits of \emph{VeriQBench} are used there. A summary of these applications is presented in Table~\ref{table:application}. 
\begin{table}[t!]
\centering
\caption{Example Applications}
\vspace{2mm}
\renewcommand{\arraystretch}{1.3}
\setlength{\tabcolsep}{1.5mm}{
\begin{tabular}{c|c}
\hline 
Example Applications & Used \emph{VeriQBench} Benchmark Circuits  \\
\hline
Equivalence Checking~\cite{burgholzer2020improved,hong2021equivalence} & QFT, Grover, Dynamic (Section \ref{sec:dynamic})\\

Fault Simulation & BV, QFT\\


Circuit Optimization~\cite{nam2018automated,amy2014polynomial} & QFT, Reversible (Section \ref{sec-reversible}) \\

ATPG~\cite{bera2017detection,chen2022automatic} & Grover, BV, QFT, QV\\

Qubit Mapping~\cite{tan2020optimality,tan2020optimal} & Generated Circuits (Section \ref{sec-qubitmapping})\\


Model Checking~\cite{gay2005probabilistic} & Teleportation\\

\hline
\end{tabular}}
\label{table:application}
\end{table}
\begin{itemize}
    \item \textit{Equivalence Checking of Quantum Circuits}: Equivalence checking is to check if two quantum circuits are functionally equivalent \cite{viamontes2007checking,yamashita2010fast,burgholzer2020improved,burgholzer2020advanced,hong2021equivalence,hong2021approximate,WLY21}. The circuits for quantum Fourier transform and Grover's algorithms have been used in work  \cite{burgholzer2020improved} to demonstrate the effectiveness of their equivalence checking methods. Also, the  circuits given in Section \ref{sec:dynamic} have been used in testing the algorithm developed in \cite{hong2021equivalence} for checking equivalence of dynamic quantum circuits, and the circuits such as Bernstein-Vazirani and quantum Fourier transform and quantum volume have been used in the approximate equivalence checking of noisy quantum circuits \cite{hong2021approximate}. 
    The circuits given in Section \ref{sec:sequential} have been used in equivalence checking of sequential quantum circuits \cite{WLY21}. 
    
    \item \textit{Fault Simulation of Quantum Circuits}: Given a quantum circuit and its faulty implementation, fault simulation is to calculate the fidelity between the expected output state and the actually obtained output state \cite{LANVQA,MPDO,SQCANC,FS}. The work is performed on combinational quantum circuits, including Bernstein-Vazirani and Quantum Fourier Transform with realistic fault models proposed in \cite{9218573}.
    \item \textit{Circuit Optimization}: Quantum circuit optimization~\cite{maslov2008quantum,prasad2006data,saeedi2013synthesis,amy2014polynomial,nam2018automated} aims to reduce the complexity of a quantum circuit, where the complexity, depending on the scenario, can be quantified by the size, depth, T-count or T-depth of the quantum circuit. The reversible circuits such as divisibility checkers have been used to benchmark the quantum circuit optimization algorithms~\cite{nam2018automated,amy2014polynomial}.
    \item \textit{Automatic Test Pattern Generation (ATPG)}: Quantum ATPG algorithms \cite{paler2012detection,bera2017detection,chen2022automatic} aim to generate specific test patterns based on the structure and fault model of the quantum circuit,
    where the test patterns include a set of input vectors and corresponding outputs vectors. The Combinational Quantum Circuits including Grover's algorithm, Bernstein-Vazirani, Quantum Fourier Transform and Quantum Volume with the unitary fault model have been used to demonstrate the effectiveness of the ATPG algorithms in \cite{bera2017detection,chen2022automatic}.
    \item \textit{Qubit Mapping}: The target of qubit mapping is to assign a physical qubit on a quantum chip to every logical qubit in a quantum circuit, while optimizing the performance of the circuit  \cite{maslov2008quantum,shafaei2014qubit,lin2014paqcs,wille2014optimal,pedram2016layout,siraichi2018qubit,zulehner2018efficient,childs2019circuit,wille2019mapping,li2019tackling,bhattacharjee2019muqut,murali2019full,tannu2019not,sivarajah2020t}. One of the most important performance metrics of quantum circuits is the depth of the circuit. The method for generating qubit mapping problems with optimal solutions for depth in this benchmark was proposed in \cite{tan2020optimality}, and was used to evaluate the qubit mapping tools in \cite{tan2020optimal}.  
    \item \textit{Model Checking Quantum Systems}: Model-checking is one of the most successful verification techniques with numerous applications in hardware and software industries. It has been generalised to check various properties of quantum systems, including quantum cryptographic protocols \cite{gay2008qmc}, quantum programs \cite{yuan2021model} and quantum circuits \cite{ying2021model}.  The teleportation circuit in Section \ref{teleportaion} has been used in probabilistic model checking algorithm of quantum protocols \cite{gay2005probabilistic}.

    
    
    
    
    
\end{itemize}

\section*{Acknowledgments}
We would like to thank Zhicheng Zhang, Minbo Gao and Riling Li for their insightful discussions. This work was partly supported by the National Key R\&D Program of China (Grant No: 2018YFA0306701), the National Natural Science Foundation of China (Grant No: 61832015). 
\sloppy
\printbibliography
\end{document}